\documentclass[twocolumn,amsmath,amssymb,aps,pra,floatfix,nofootinbib,superscriptaddress]{revtex4}

\usepackage{bm} \usepackage{graphicx} 
\usepackage{amsmath}
\usepackage{amsthm,stmaryrd,mathtools}
\usepackage{amssymb} 
\usepackage{epstopdf} 
\usepackage{txfonts}
\usepackage{graphicx}
\usepackage{siunitx}
\usepackage{hyperref}
\usepackage{braket}
\usepackage{amsmath}
\DeclareMathOperator{\Tr}{Tr}

\usepackage{graphicx,graphics,times,color}
\usepackage{bm}
\usepackage{longtable}
\usepackage{color}
\usepackage{lipsum}
\usepackage{graphicx}
\usepackage{epstopdf}

\def\be{\begin{equation}}
\def\ee{\end{equation}}
\def\ba{\begin{eqnarray}}
\def\ea{\end{eqnarray}}

\begin{document}

\title{Certification of spin-based quantum simulators }

\author{Abolfazl Bayat}
\affiliation{Institute of Fundamental and Frontier Sciences, University of Electronic Science and Technology of China, Chengdu 610051, China}
\affiliation{Department of Physics and Astronomy, University College London, London WC1E 6BT, United Kingdom}

\author{Benoit Voisin}
\affiliation{Centre for Quantum Computation and Communication Technology, School of Physics, The University of New South Wales, Sydney, New South Wales 2052, Australia}

\author{Gilles Buchs}
\affiliation{Centre for Quantum Computation and Communication Technology, School of Physics, The University of New South Wales, Sydney, New South Wales 2052, Australia}

\author{Joe Salfi}
\affiliation{Centre for Quantum Computation and Communication Technology, School of Physics, The University of New South Wales, Sydney, New South Wales 2052, Australia}
\affiliation{Department of Electrical and Computer Engineering, University of British Columbia, Vancouver, BC V6T 1Z4, Canada}

\author{Sven Rogge}
\affiliation{Centre for Quantum Computation and Communication Technology, School of Physics, The University of New South Wales, Sydney, New South Wales 2052, Australia}

\author{Sougato Bose}
\affiliation{Department of Physics and Astronomy, University College London, London WC1E 6BT, United Kingdom}

\date{\today}

\begin{abstract}
Quantum simulators are engineered devices controllably designed to emulate complex and classically intractable quantum systems. A key challenge is to certify whether the simulator truly mimics the Hamiltonian of interest. This certification step requires the comparison of a simulator's output to a known answer, which is usually limited to small systems due to the exponential scaling of the Hilbert space. Here, in the context of Fermi-Hubbard spin-based analogue simulators, we propose a modular many-body spin to charge conversion scheme that scales linearly with both the system size and the number of low-energy eigenstates to discriminate. Our protocol is based on the global charge state measurement of a 1D spin chain performed at different detuning potentials along the chain. In the context of semiconductor-based systems, we identify realistic conditions for detuning the chain adiabatically in order to avoid state mixing while preserving charge coherence. Large simulators with vanishing energy gaps, including 2D arrays, can be certified block-by-block with a number of measurements scaling only linearly with the system size.  
\end{abstract}


\maketitle

\section{Introduction} 

Quantum simulators~\cite{georgescu2014quantum,cirac2012goals,buluta2009quantum,bernien2017probing,zhang2017observation,dutta2012nonequilibrium,barends2014superconducting,o2016scalable,roushan2017spectroscopic,barends2015digital} are devices designed to emulate the behavior of quantum systems in order to provide new insights into complex quantum phenomena~\cite{anderson1987resonating,lee2006doping,banerjee2013atomic,zhang2018experimental,PhysRevA.94.063626,daley2012measuring,trotzky2012probing,lv2018quantum}, solving complex optimization problems~\cite{pagano2019quantum,ding2019towards} or realizing models that do not exist naturally~\cite{kitaev2003fault}. Analog quantum simulators, which evolve according to a designed global Hamiltonian, offer an efficient way to implement some of these problems without the need for a complete set of local gates and readouts mechanisms. One of the main challenges is to certify that a large-scale quantum simulator, non-tractable classically due to the exponential divergence of the Hilbert space with the system size, truly emulates the task that it is designed for~\cite{wiebe2014hamiltonian,gao2017quantum,hangleiter2017direct,aolita2015reliable,Arute2019}. 
However, many certification protocols rely on full quantum state tomography~\cite{PhysRevLett.105.150401}, which requires local addressability, and results in a number of measurements that diverges exponentially with the system size. Variational methods~\cite{Kokail2019} can scale more favorably, but they still require local addressability and implementing classical optimization schemes becomes challenging as the system size increases. Hence, these techniques are not suitable for large-scale analogue quantum simulators and new certification schemes are highly sought for.

\begin{figure} \centering
	\includegraphics[width=8cm,height=4cm,angle=0]{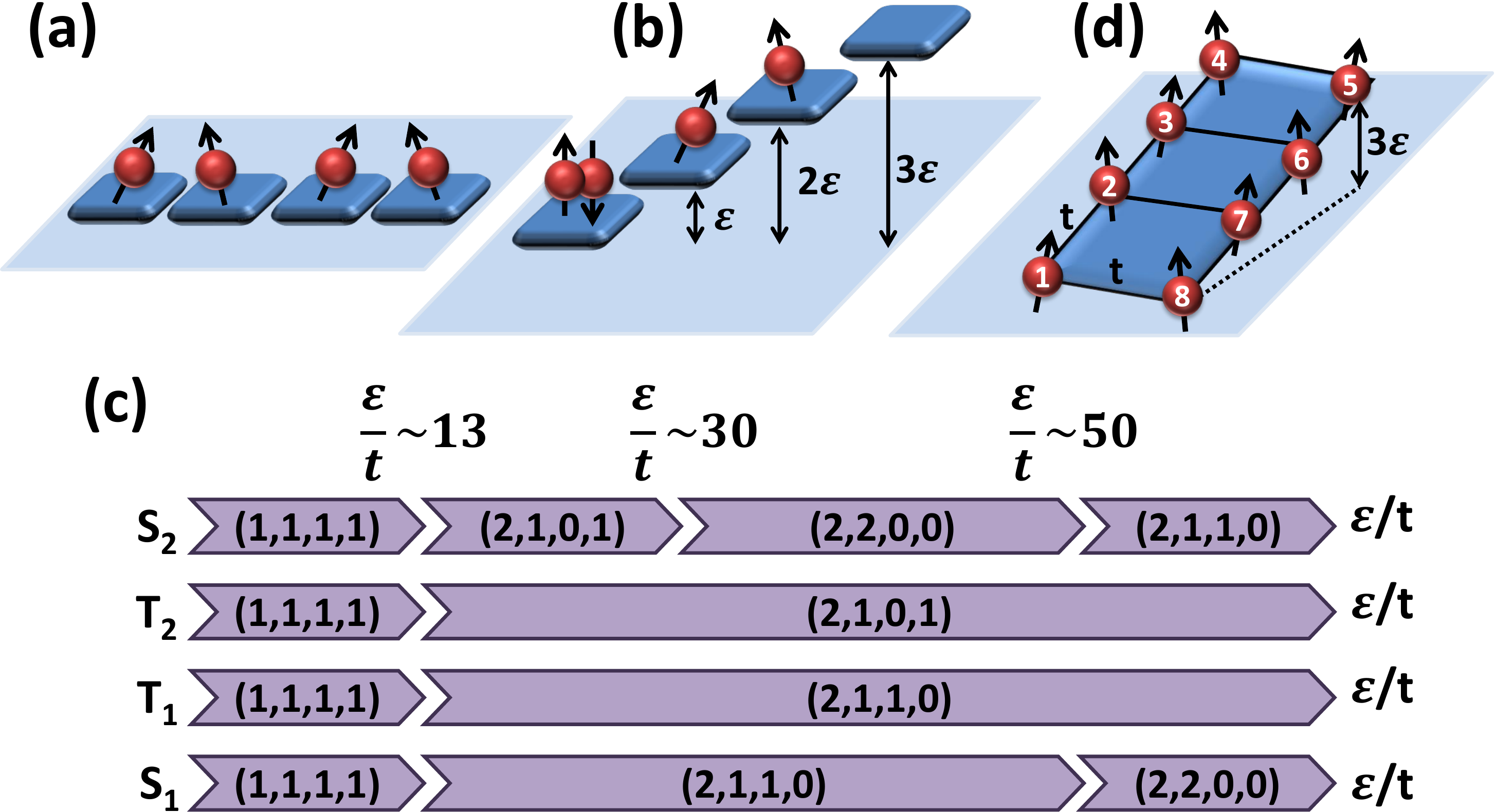}
	\caption{ \textbf{Schematic of the system.} (a) Chain of interacting electrons confined in the sites of a regular lattice. (b) Potential gradient (tilt) applied across the chain aiming at a change in the charge configuration. (c) Low energy spectrum and corresponding charge configuration in the tilted system. (d) Geometry of a spin ladder of size $2 \times 4$ sites, where tunneling and Coulomb interactions between neighboring sites in both vertical and horizontal directions are $t$ and $V$ respectively.}
	\label{Fig_Schematic}
\end{figure} 

Here, in the context of semiconductor-based systems, we propose a spin-to-charge conversion readout scheme able to discriminate between the low energy entangled spin eigenstates of a Heisenberg spin array. The Heisenberg model is a key model in condensed matter physics~\cite{sachdev2011quantum,amico2008entanglement}, spintronics~\cite{vzutic2004spintronics} and quantum technologies~\cite{farooq2015adiabatic,yang2010spin}. Our certification protocol relies on global charge configuration measurements of the simulator under different potential gradients (called tilts) applied adiabatically along a spin chain. The strength of our certification protocol is the ability to discriminate between eigenstates sharing the same symmetries and total spin (differing only in their entanglement structure) without local spin addressing. Using global over local measurements is here the key to develop a protocol that scales favorably. We demonstrate how chains of $N\le 10$ spins can be certified with high fidelity in the required tractable regime. These finite chains constitute the building blocks to extend our certification protocol to larger systems, including 2D arrays. Here, a modular approach is developed. The simulator is certified block-by-block, resulting in a number of measurements which scales linearly with the system size as well as with the number of eigenstates to discriminate.

\section{Model} 

To study the Heisenberg model we consider $N$ electrons hopping among $N$ sites (half filling) in a 1D lattice, following the Fermi-Hubbard model:
\begin{eqnarray} \label{Hubbard_Hamiltonian}
H &=& t \sum_{\langle k,l \rangle}  \sum_{\sigma=\uparrow,\downarrow}  \left( c_{k,\sigma}^\dagger c_{l,\sigma}+ c_{l,\sigma}^\dagger c_{k,\sigma} \right) + \sum_{k=1}^{N} \tilde{\epsilon}_k n_k \cr 
&+& V \sum_{\langle k,l \rangle} n_k n_{l} + \frac{U}{2} \sum_{k=1}^{N} n_k (n_k-1),
\end{eqnarray} 
where $c_{k,\sigma}$ ($c_{k,\sigma}^\dagger$) is the annihilation (creation) fermionic operator for an electron at site $k$ with spin $\sigma$, number operator $n_k=\sum_{\sigma=\uparrow,\downarrow}c_{k,\sigma}^\dagger c_{k,\sigma}$ counts the number of electrons at site $k$, $t$ is the tunnel coupling between neighboring sites, $\tilde{\epsilon}_k$ is the local potential at site $k$, $V$ is the Coulomb interaction between adjacent sites, and $U$ is the on-site energy and $\langle k,l \rangle$ means summation over the nearest neighbor sites $k$ and $l$. 
In the case of a homogeneous 1D array, i.e. $\tilde{\epsilon}_k{=}0$,  and without Coulomb interaction, i.e. $V=0$,  the Hamiltonian (\ref{Hubbard_Hamiltonian}) is solvable~\cite{Elliott1968}. 
Throughout this paper we consider a chain made of an even number of sites $N$, with on-site energy $U/t{=}40$, Coulomb interaction $V/t{=}10$. These values are chosen to match an experimental situation detailed in section~\ref{sec:Experimental_Realization}. The local potential is of the form 
$\tilde{\epsilon}_k=(k-1)\epsilon$ where $\epsilon$ is the potential difference between two adjacent sites.
A schematic of the system is shown in Fig.~\ref{Fig_Schematic}(a).
Note that the Hamiltonian $H$ commutes with the total number of electrons $\sum_k n_k$, for all values of $\epsilon$. Thus, the filling factor is a conserved quantity. 
In a homogeneous lattice ($\tilde{\epsilon}_k{=}0$), whenever $U{\gg} t$, the low energy eigenstates take the charge configuration $(1,1,\cdots,1)$ and the system effectively becomes a Heisenberg spin chain with exchange coupling $J{\sim} t^2/U$ (with possible corrections due to $V$)~\cite{pica2014exchange}. These eigenstates form a low energy manifold separated by units of $U$ from the eigenstates with double charge occupancies for which the map to the Heisenberg model fails. 
For even $N$ the ground state $|S_1\rangle$ is always a global singlet with total spin $S_{tot}{=}0$. The first two excited states $|T_1\rangle$ and $|T_2\rangle$ are triplets with the total spin $S_{tot}{=}1$. 
The fourth eigenstate is again another global singlet $|S_2\rangle$. In a chain of length $N=4$ these four eigenstates form the low energy manifold. 
It is worth emphasizing that the total spin is also a conserved quantity of the Hamiltonian $H$. This means that even at finite local detuning $\epsilon$ each eigenvector of the system always conserves its total spin.    
 

\begin{figure} \centering
	\includegraphics[width=7.5cm,height=5.5cm,angle=0]{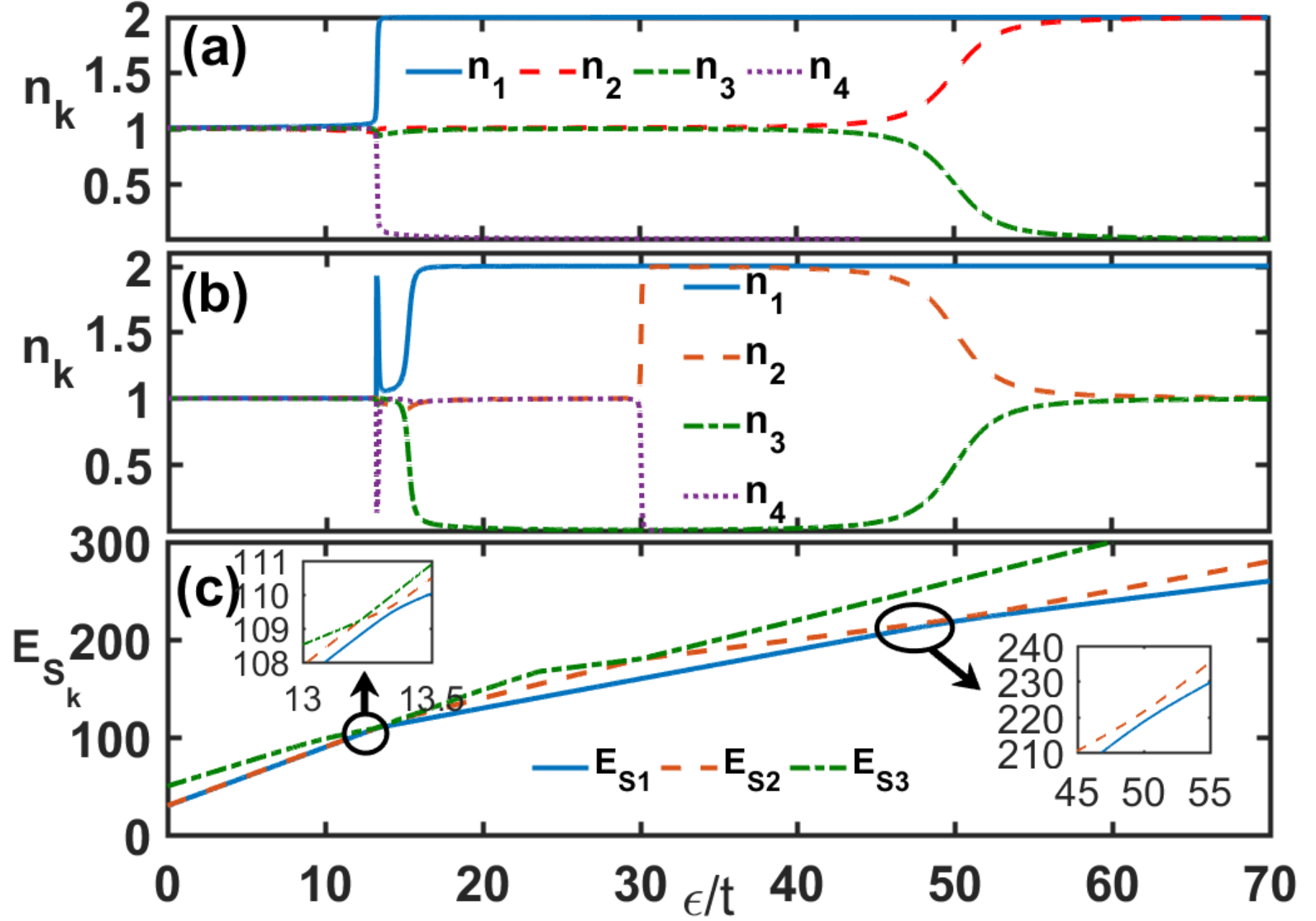} 
	\caption{ \textbf{Singlet charge configurations.} Charge occupancies of a chain of length $N=4$ for the state: (a) $|S_1\rangle$; and (b) $|S_2\rangle$. (c) Energy spectrum of the first three singlet eigenstates.  }
	\label{Fig_Charge_S_N4}
\end{figure}

\section{Charge configurations} 

Many-body spin eigenstate measurement is a challenging task. For example, $|S_1\rangle$ and $|S_2\rangle$ have the same total spin $S_{tot}{=}0$ and share various symmetries (e.g. SU(2) invariance) making them difficult to be distinguished locally. To achieve spin eigenstate readout, we apply a potential tilt across the chain, i.e. a finite $\epsilon$, to provide enough energy for electrons to overcome $U$, as shown in Fig.~\ref{Fig_Schematic}(b), and then the charge configuration is measured. 
Since the eigenstates are always orthogonal, their experimentally measurable charge configurations depend on their spin state. This constitutes  the core of our certification method.

\begin{figure} \centering
	\includegraphics[width=7.5cm,height=6cm,angle=0]{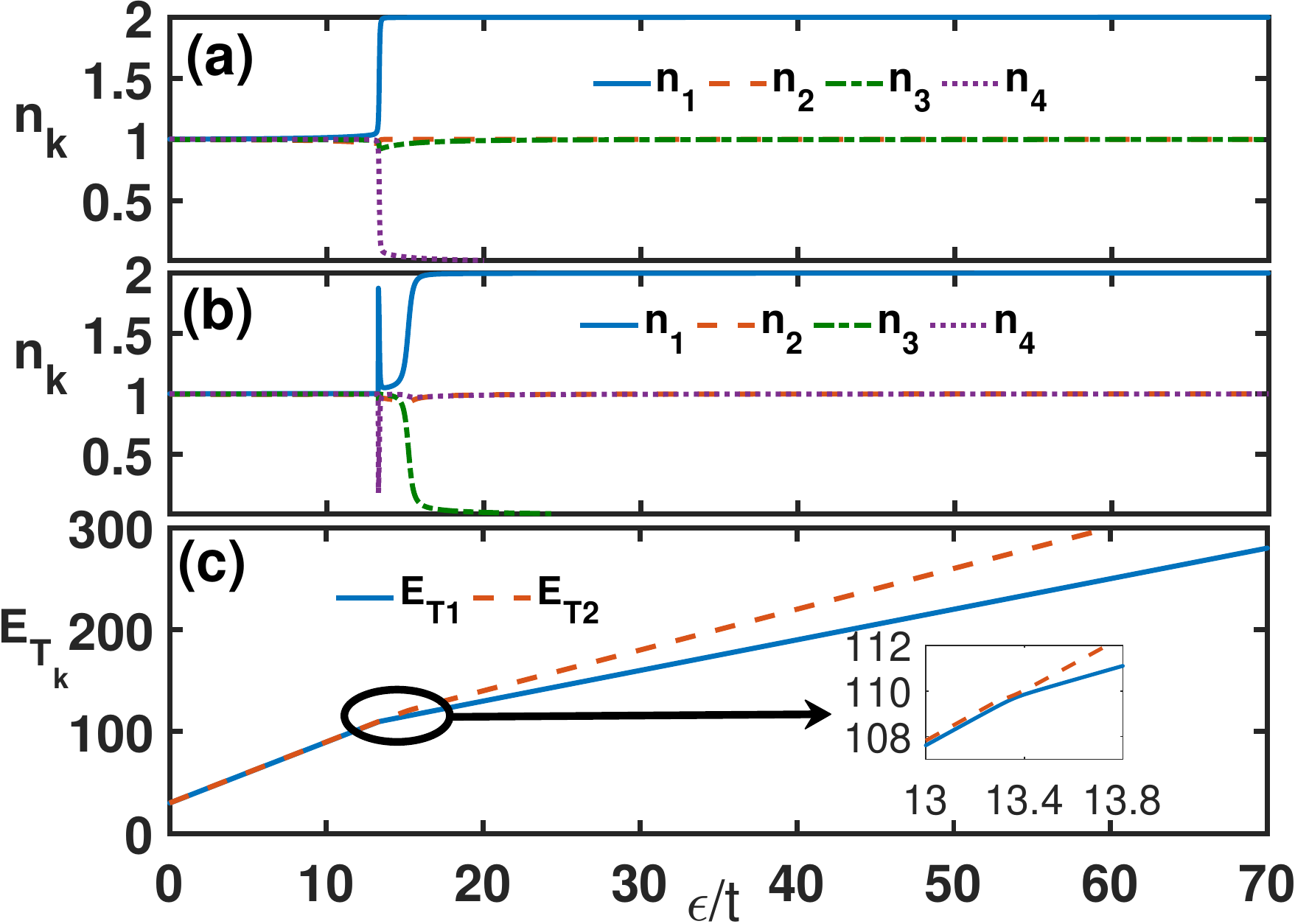} 
	\caption{ \textbf{Triplet charge configurations.} Charge occupancies of a chain of length $N=4$ for the state: (a) $|T_1\rangle$; and (b) $|T_2\rangle$. (c) Energy spectrum of the first two triplet eigenstates. }
	\label{Fig_Charge_T_N4}
\end{figure}

We now develop the evolution of the charge configurations versus tilt for a chain of $N{=}4$. Longer chains are discussed in section~\ref{section_large_chain}. The charge configurations of the two singlet eigenstates $|S_1\rangle$ and $|S_2\rangle$ as a function of $\epsilon/t$ are plotted in Figs.~\ref{Fig_Charge_S_N4}(a)-(b). 
The charge configuration changes for both eigenstates around $\epsilon/t {\sim} 13.4$ and one electron moves from either site $4$ (in the case of $|S_1\rangle$) or site $3$ (in the case of $|S_2\rangle$) to site $1$, creating two different charge configurations for $|S_1\rangle$ and $|S_2\rangle$. At around $\epsilon/t{\sim} 30$ in the eigenstate $|S_2\rangle$ an electron moves from site $4$ to site $2$ resulting in the charge configuration $(2,2,0,0)$. Finally, at $\epsilon/t {\sim} 50$ the charge configuration of $|S_2\rangle$ evolves to $(2,1,1,0)$ while $|S_1\rangle$ rearranges to $(2,2,0,0)$. All these charge configurations are summarized in Fig.~\ref{Fig_Schematic}(c). 
To understand the charge dynamics we plot the energies of the first three singlet eigenstates in Fig.~\ref{Fig_Charge_S_N4}(c). Any charge movement in the eigenstates corresponds to an anti-crossing between two eigenstates with the same $S_{tot}$. This is evident at $\epsilon/t {\sim} 13.4$,  $\epsilon/t {\sim} 30$ and $\epsilon/t {\sim} 50$ where $E_{S_1}$ and $E_{S_2}$, $E_{S_2}$ and $E_{S_3}$ and $E_{S_1}$ and $E_{S_2}$ again anti-cross, respectively. 
Interestingly, the charge occupancy spike which occurs in $|S_2\rangle$ at $\epsilon/t\sim 13.4$ just before the first charge movement (see Fig.~\ref{Fig_Charge_S_N4}(b)), can also be associated to an anti-crossing, between the $|S_2\rangle$ and $|S_3\rangle$ states (see inset of Fig.~\ref{Fig_Charge_S_N4}(c)).


A similar analysis can be performed for the triplet states. The charge configurations of the two triplets $|T_1\rangle$ and $|T_2\rangle$ are depicted in Figs.~\ref{Fig_Charge_T_N4}(a)-(b), respectively. 
The charge configuration of both eigenstates changes around $\epsilon/t {\sim} 13.4$ and one electron moves from either site $4$ (in the case of $|T_1\rangle$) or site $3$ (in the case of $|T_2\rangle$) to site $1$. 
In Fig.~\ref{Fig_Charge_T_N4}(c) we plot the energy eigenvalues of both  $|T_1\rangle$ and $|T_2\rangle$ as functions of $\epsilon/t$ showing an anti-crossing at the charge transition point $\epsilon/t {\sim} 13.4$. 
As we will see later, even for larger systems, the final charge configurations are always $(2,\cdots,2,0,\cdots,0)$ for $|S_1\rangle$ and 
$(2,\cdots,2,1,1,0,\cdots,0)$ for $|T_1\rangle$. We will show that this important feature can be used for certification.


\section{States discrimination} 

First we consider the ideal case in which the potential tilting is performed adiabatically for all eigenstates. 
The number of required tilts depends on the number of eigenstates to be discriminated. For instance, to distinguish between $|S_1\rangle$  and  $|T_1\rangle$ only one tilt, namely $\epsilon/t\simeq 50-60$, is needed as $|S_1\rangle$ takes the configuration $(2,2,0,0)$ and $|T_1\rangle$ goes to $(2,1,1,0)$. 
Only two tilts are required to fully distinguish the four lowest eigenstates. For instance, by tilting to $\epsilon/t{=}35$, $|S_2\rangle$ can be fully distinguished, with configuration $(2,2,0,0)$, and  $|T_2\rangle$, with configuration $(2,1,0,1)$. However, both  $|S_1\rangle$  and  $|T_1\rangle$ share the same configuration $(2,1,1,0)$ and cannot be distinguished. Therefore, another charge configuration measurement must be performed at a larger detuning 
$\epsilon/t {\sim} 50-60$ when the charge configuration for $|S_1\rangle$ changes to $(2,2,0,0)$ while $|T_1\rangle$ remains in the $(2,1,1,0)$ configuration.
The key feature of our proposal lies in its scalability: only two tilts are needed to fully distinguish the four lowest eigenstates, irrespective of the system size (see the section~\ref{section_large_chain}). In fact, for distinguishing $n$ low-energy eigenstates, only $n/2$ tilts are required.

\begin{figure} \centering
	\includegraphics[width=7.5cm,height=6cm,angle=0]{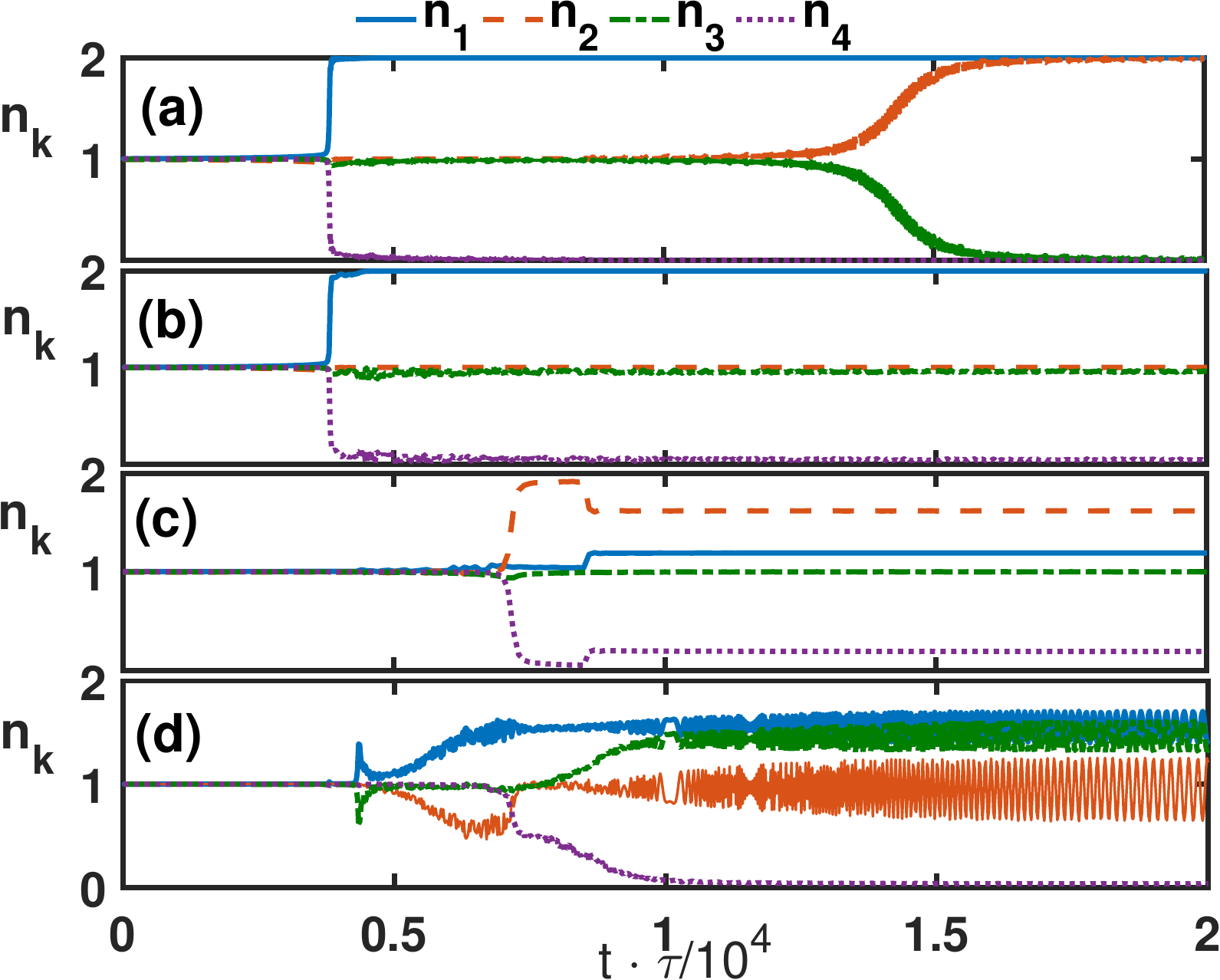}
	\caption{ \textbf{Adiabatic evolution.} Charge occupancies in the evolution of a system of length $N=4$ when $T_{max}{=}2\times 10^4/t$ and the system is initialized in the state: (a) $|S_1\rangle$; (b) $|T_1\rangle$; (c) $|T_2\rangle$; and (d) $|S_2\rangle$. 
		This choice of $T_{max}$ results in an adiabatic evolution only for $|S_1\rangle$ and $|T_1\rangle$.	}
	\label{Fig_Adiabatic_ST_N4_all}
\end{figure}

\section{Adiabatic Evolution}

In a practical scenario, in order to readout the many-body spin eigenstate, we tilt the system, initially prepared in one of the low energy eigenstates, adiabatically such that it remains in the local eigenvector of the Hamiltonian at any time $\tau$. 
The eigenstates can be discriminated by measuring the charge configuration at different potentials $\epsilon$. The tilt potential varies as 
\begin{equation}\label{epsilon_tau}
\epsilon (\tau)=
\left\{ {\begin{array}{c}
	\frac{\tau}{T_{max}}  \epsilon_{max},  \quad \text{ for: } \tau \le T_{max}  \\
	\epsilon_{max}, \quad \qquad \text{for: } \tau > T_{max}   \\    
	\end{array}  } \right.
\end{equation}
where $\epsilon_{max}$ is the maximum tilt potential considered here to be $\epsilon_{max}/t=70$. For any initial state $|\Psi(0)\rangle$ the system evolves to the state $|\Psi(\tau)\rangle$ according to the Schr\"{o}dinger equation under the action of the time dependent Fermi-Hubbard Hamiltonian described in Eq.~(\ref{Hubbard_Hamiltonian}). 
The choice of $T_{max}$ is important as it results in different system dynamics.
Adiabaticity, which notably protects the evolution against Landau-Zener transitions while sweeping through anticrossings, is achieved for slow dynamics and large $T_{max}$. 
An upper bound for $T_{max}$ is however set by the coherence time of the system, as decoherence occurs at the charge transitions which are swept through. In Fig.~\ref{Fig_Adiabatic_ST_N4_all}(a) we plot the charge occupancies for the quantum state $|\Psi(\tau)\rangle$, taking $T_{max}=2\times 10^4/t$, as a function of time when the system is initially prepared in the state $|S_1\rangle$. The charge configurations are very similar to the real eigenstates displayed in Fig.~\ref{Fig_Charge_S_N4}(a), with the fidelity of the evolution  $F=\left| \langle \Psi(\tau)|S_1(\tau)\rangle \right|^2$ remaining above 0.98 throughout the evolution, which demonstrates that the adiabatic condition is well satisfied. In Fig.~\ref{Fig_Adiabatic_ST_N4_all}(b) we depict the charge occupancies when the system is initialized in the state $|T_1\rangle$. Again the charge configurations are very similar to the ones for the real eigenstate shown in Fig.~\ref{Fig_Charge_T_N4}(a) with the fidelity above $0.97$ throughout the evolution. In Figs.~\ref{Fig_Adiabatic_ST_N4_all}(c) and (d) we plot the charge occupancies of the state $|\Psi(\tau)\rangle$ when the system is initially in the state $|T_2\rangle$ and $|S_2\rangle$, respectively. In these two cases, the evolution is very different from the charge configurations of the local eigenstates given in Fig.~\ref{Fig_Charge_T_N4}(b) and  Fig.~\ref{Fig_Charge_S_N4}(b), respectively. Here $T_{max}$ is not large enough to keep an adiabatic evolution for these two eigenstates and their fidelity reaches levels  as low as ${\sim} 0.2$. 
In fact, to make the evolution adiabatic for $|S_2\rangle$ and $|T_2\rangle$ one has to take $T_{max}$ to be $(10^7-10^8)/t$ due to smaller gaps between higher energy eigenstates. 
For instance, in a chain of length $N=4$, the energy gaps between the first two singlet and triplet states are $(E_{S_2}-E_{S_1})/t=0.2231$  and $(E_{T_2}-E_{T_1})/t=0.0913$, respectively. However, these gaps become smaller for higher states as $(E_{S_3}-E_{S_2})/t= 0.0533$  and $(E_{T_3}-E_{T_2})/t= 0.0238$, respectively.

Remarkably, only an adiabatic evolution of  $|S_1\rangle$ and $|T_1\rangle$ is enough to distinguish all four eigenstates, enabling complete certification. Let's consider an evolution which is only adiabatic for $|S_1\rangle$ and $|T_1\rangle$, like depicted in Fig.~\ref{Fig_Adiabatic_ST_N4_all}. For $|S_2\rangle$, the outcome of the charge measurement will be time averaged over the charge occupancies due to rapid charge oscillations. Therefore, by using the same procedure, at $\epsilon/t{=}35$ the states $|S_2\rangle$ and $|T_2\rangle$ take the configurations  $(1.5,1,1.5,0)$ and $(1.2,1.6,1,0.2)$, respectively, which are very distinct from each other as well as from the configuration  of $|S_1\rangle$ and $|T_1\rangle$. Note that the partial charges mean that the quantum states are in a  superposition of multiple charge states.  This means that even when the evolution for $|S_2\rangle$ and $|T_2\rangle$ is  non-adiabatic the proposed discrimination procedure still holds.

\begin{table}[h!] 
	\centering
	\begin{tabular}{ | m{1cm} | m{1cm}| m{1cm} | m{1cm}| m{1cm} | } 
		\hline
		N & 4 & 6 & 8 & 10 \\ 
		\hline
		$\Delta E_{S}^{(21)}/t$ & 0.2231 &  0.1255 &  0.1011 & 0.0710 \\ 
		\hline
		$\Delta E_{T}^{(21)}/t$ & 0.0913 & 0.0757 & 0.0684 & 0.0575 \\ 
		\hline
	\end{tabular}
	\caption{ Minimum energy gaps for both singlet and triplet states during the evolution of the system for $\epsilon_{max}/t$ varies from $0$ to $70$.   } 
	\label{Table_Energy_Gap}
\end{table}

A crucial issue for the adiabatic evolution is the estimation of $T_{max}$ needed to evolve larger systems. As we discussed above, it is important to keep the evolution for both $|S_1\rangle$ and $T_1\rangle$ adiabatic, even if the higher energy eigenstates do not follow an adiabatic evolution. The criteria for the validity of the adiabatic theorem has been a subject of research for many years~\cite{comparat2009general,marzlin2004inconsistency,tong2007sufficiency}. A standard criterion implies that one has to satisfy $|\frac{\langle \dot{S}_2(\tau)|S_1(\tau)\rangle }{E_{S_2}(\tau)-E_{S_1}(\tau)}|\ll 1$ where $|\dot{S}_2(\tau)\rangle$ is the time derivative of the eigenstate $|S_2\rangle$ with respect to $\tau$. A similar criterion can be written for triplets as well. Using perturbation theory one can show that in a pessimistic estimation $\langle \dot{S}_2(\tau)|S_1(\tau)\rangle \sim T_{max}^{-1} [ E_{S_2}(\tau)-E_{S_1}(\tau)]^{-1}$. This implies that for the validity of the adiabatic evolution one has to keep $T_{max}>1/\Delta E^2$  where $\Delta E$ is the the energy gap. To see how the energy gap scales with system size we present the minimum energy gap during the adiabatic evolution for both singlets (i.e. $\Delta E_S=E_{S_2}-E_{S_1}$) and triplets (i.e. $\Delta E_T=E_{T_2}-E_{T_1}$) in TABLE~\ref{Table_Energy_Gap}. As the data show, the energy gap decreases fairly linearly as the system size increases. This means that for a chain of size $N=10$ the time $T_{max}$ is almost $10$ times larger than the one needed for $N=4$.

\begin{figure} \centering
	\includegraphics[width=7.5cm,height=5.5cm,angle=0]{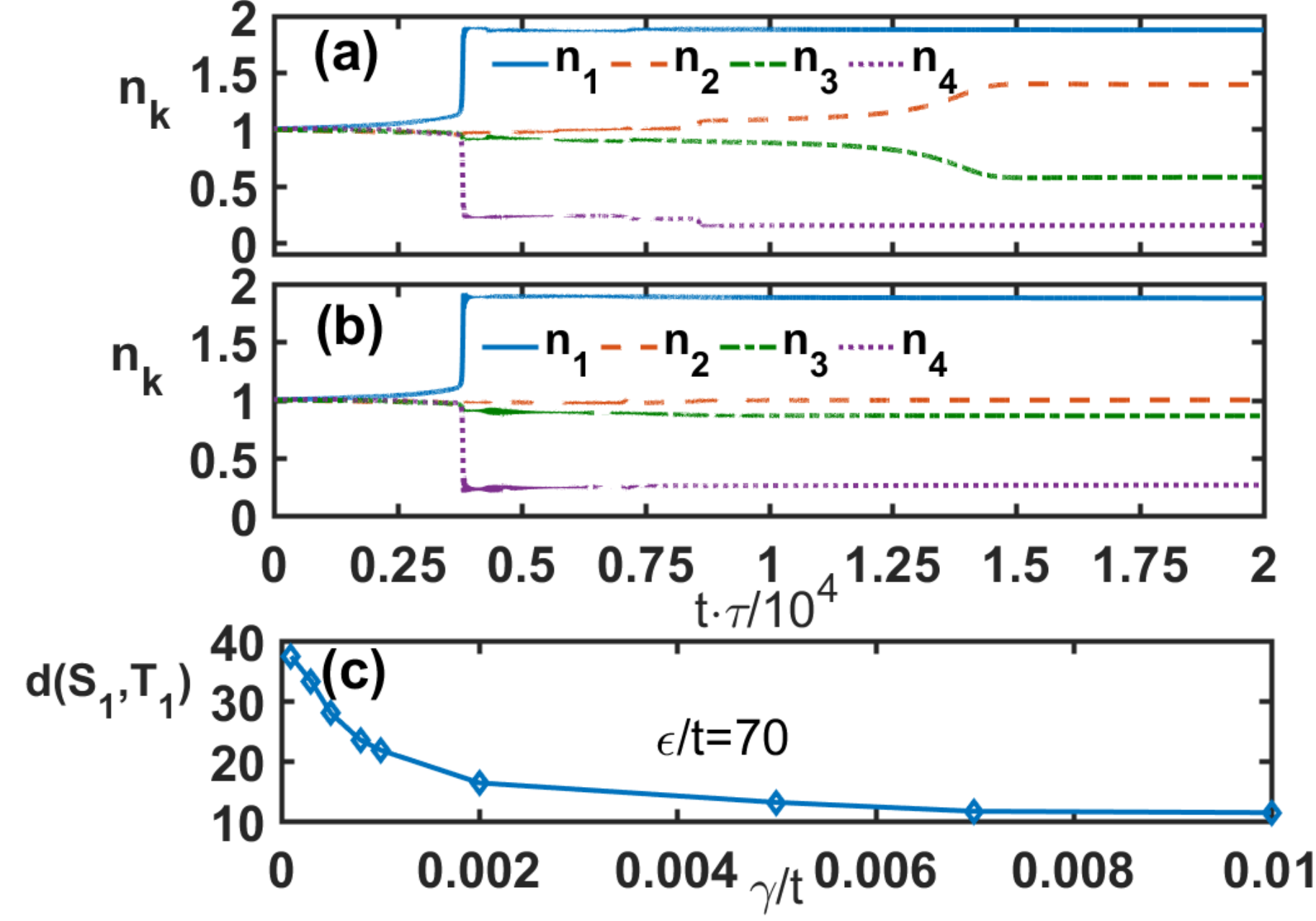}
	\caption{ \textbf{Decoherence.} Time evolution in the presence of decoherence in a system of length $N=4$ when $T_{max}{=}2\times 10^4/t$. Charge occupancies are given for $\gamma/t=0.001$ when the system is initialized in the eigenstate: (a) $|S_1\rangle$; and (b) $|T_1\rangle$. (c) Distance between the charge probability distributions of $|S_1\rangle$ and $|T_1\rangle$ as a function of $\gamma$ when $\epsilon/t=70$.}  
	\label{Fig_Decoherence_ST_N4}
\end{figure}

\begin{figure} \centering
	\includegraphics[width=8cm,height=7cm,angle=0]{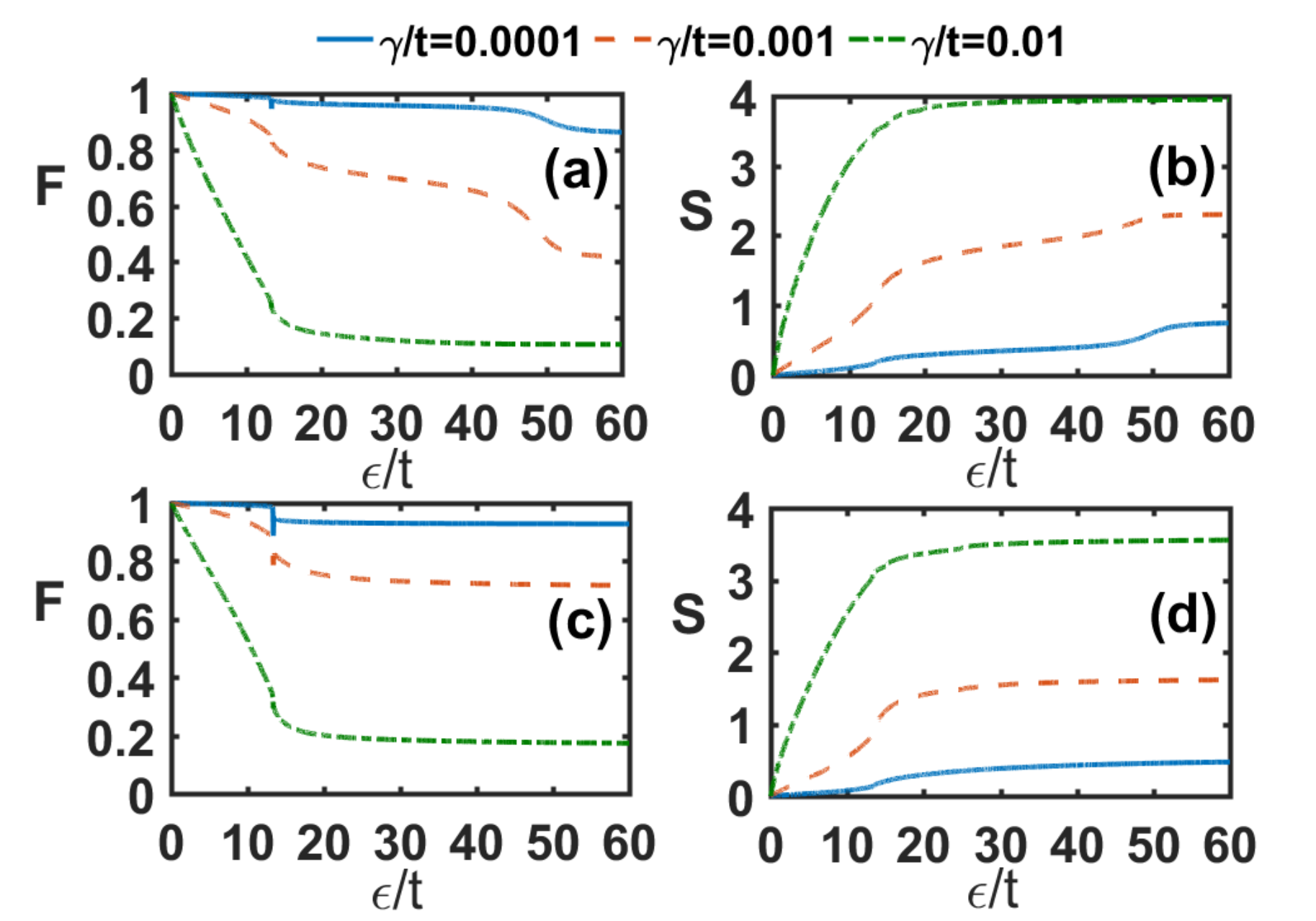}
	\caption{ \textbf{Fidelity and Entropy.} Evolution of the fidelity and the entropy in a system of length $N{=}4$ in the presence of decoherence. For convenience, the quantities are plotted as a function of $\epsilon/t$ for which $\epsilon$ is given in Eq.~(\ref{epsilon_tau}). Panels represent: (a) The fidelity $F$ for $|S_1\rangle$. (b) The von Neumann entropy $S$ for $|S_1\rangle$. (c) The fidelity $F$ for $|T_1\rangle$. (d) The von Neumann entropy $S$ for $|T_1\rangle$.  }
	\label{Fig_Fidelity_Entropy_Decoherence_ST_N4}
\end{figure}

\section{Decoherence}  

Interaction with the environment results in non-unitary dynamics and decoherence. For itinerant particles, charge fluctuations constitute the most common source of decoherence ~\cite{wardrop2014exchange,nakajima2018coherent}, leading to destruction of the superposition of different charge configurations.
Therefore, if $\{L_n\}$ represent projection operators on the $n$-th charge configuration then the dynamics is given by the Lindblad master equation
\begin{equation} \nonumber
\frac{\partial \rho}{\partial \tau}=-i[H(\tau),\rho]+\gamma \sum_n \left( L_n \rho L_n^\dagger -\frac{1}{2} L_n^\dagger L_n \rho
-\frac{1}{2} \rho L_n^\dagger L_n  \right)
\label{Lindblad_master}
\end{equation}
where $\gamma$ represents the decoherence strength, $\rho$ is the density matrix of the system and $L_n$'s are the Lindblad operators. 
Each Lindblad operator is given by a specific charge configuration that reads 
\begin{equation} 	
L_n=|n_1,n_2,\cdots,n_N\rangle \langle n_1,n_2,\cdots,n_N|
\end{equation}	
where $n_k$is the charge occupancy of site $k$. 
In Fig.~\ref{Fig_Decoherence_ST_N4}(a) we plot the charge occupancies for the evolution of $|S_1\rangle$ in a chain with $N=4$ for $\gamma/t = 10^{-3}$ and $T_{max}=2\times 10^4/t$. As the figure shows, decoherence leads to partial charge transitions and as a consequence, the quantum states become mixtures of charge configurations. The same evolution for the triplet state $|T_1\rangle$ is depicted in Fig.~\ref{Fig_Decoherence_ST_N4}(b). Its evolution is less affected than for $|S_1\rangle$ as there are less charge transitions. 

As decoherence affects charge transitions, it is important to address its impact on our protocol for distinguishing between quantum states.
Each measurement outcome is associated with a charge projection operator $L_n$ with respective probability $p_n=\Tr{(\rho L_n)}$. 
Distinguishing between the two eigenstates, e.g. $|S_1\rangle$  and $|T_1\rangle$, is equivalent to distinguishing between two probability distributions $\{p_n: p_n=\Tr{(\rho_{S_1} L_n)}\}$ and $\{q_n: q_n=\Tr{(\rho_{T_1} L_n)}\}$, where $\rho_{S_1}$ ($\rho_{T_1}$) is the solution of the above Lindblad master equation with the initial state $|S_1\rangle$  ($|T_1\rangle$). 
Experimentally, the real probability distribution can be obtained by averaging over $M$ charge measurements at each tilt.
The distance (or relative entropy) defined as $d(S_1,T_1){=}\sum_n p_n \log_2 \frac{p_n}{q_n}$ can be used to quantify the distinguishability between the two distributions. 
The error in discriminating between the two probability distributions after $M$ samples scales as ${\sim} 2^{-Md}$~\cite{vedral1997statistical}, for $M$ large. Therefore, by repeating the experiments at each tilt for $M{\sim} 10^2{-}10^3$ one can reconstruct the probability distributions and discriminate between the eigenstates when $d{>}1$.
In  Fig.~\ref{Fig_Decoherence_ST_N4}(c) we plot $d(S_1,T_1)$ versus $\gamma$ for a tilt set to $\epsilon/t{=}70$. 
The distance drops as $\gamma$ increases, however it remains above $10$ even for $\gamma/t{=}0.01$, thus discrimination is still achievable.

In order to understand the full effect of decoherence, in Fig.~\ref{Fig_Fidelity_Entropy_Decoherence_ST_N4}(a), we plot the fidelity of the evolution for the state $|S_1\rangle$ as a function of time $\tau$ for different values of noise strength $\gamma$. As the figure shows, by increasing $\gamma$ the fidelity decreases. To understand this, it is important to note that such dynamics is not unitary. This means that the quantum state of the system becomes mixed during the time evolution. To see this, one can compute the von Neumann entropy of the whole system which is defined as
\begin{equation} \label{von_Neumann_Entropy}
S(\rho)=-\Tr\left( \rho\log_2 \rho \right).
\end{equation}    
In Fig.~\ref{Fig_Fidelity_Entropy_Decoherence_ST_N4}(b) we plot the von Neumann entropy of the system when the quantum state is initially $|S_1\rangle$ as a function of time $\tau$ for different values of noise strength $\gamma$. As the figure shows the entropy increases monotonically and sharp rises happen during the charge movements when the charge state is delocalized. In Fig.~\ref{Fig_Fidelity_Entropy_Decoherence_ST_N4}(c) we also plot the fidelity for the quantum state $|T_1\rangle$ keeping all the parameters the same as for the singlet $|S_1\rangle$. Finally, in Fig.~\ref{Fig_Fidelity_Entropy_Decoherence_ST_N4}(d) we plot the von Neumann entropy of the evolution of the triplet state $|T_1\rangle$ as a function of time.  Figs.~\ref{Fig_Fidelity_Entropy_Decoherence_ST_N4}(c)-(d) show that the fidelity of the triplet is slightly higher and its von Neumann entropy is smaller in comparison with the singlet. This is due to less charge movements for triplets, or equivalently fewer energy anti-crossings between the eigenstates, which makes the triplet evolution less prone to decoherence.

\begin{figure} \centering
	\includegraphics[width=7.5cm,height=5.5cm,angle=0]{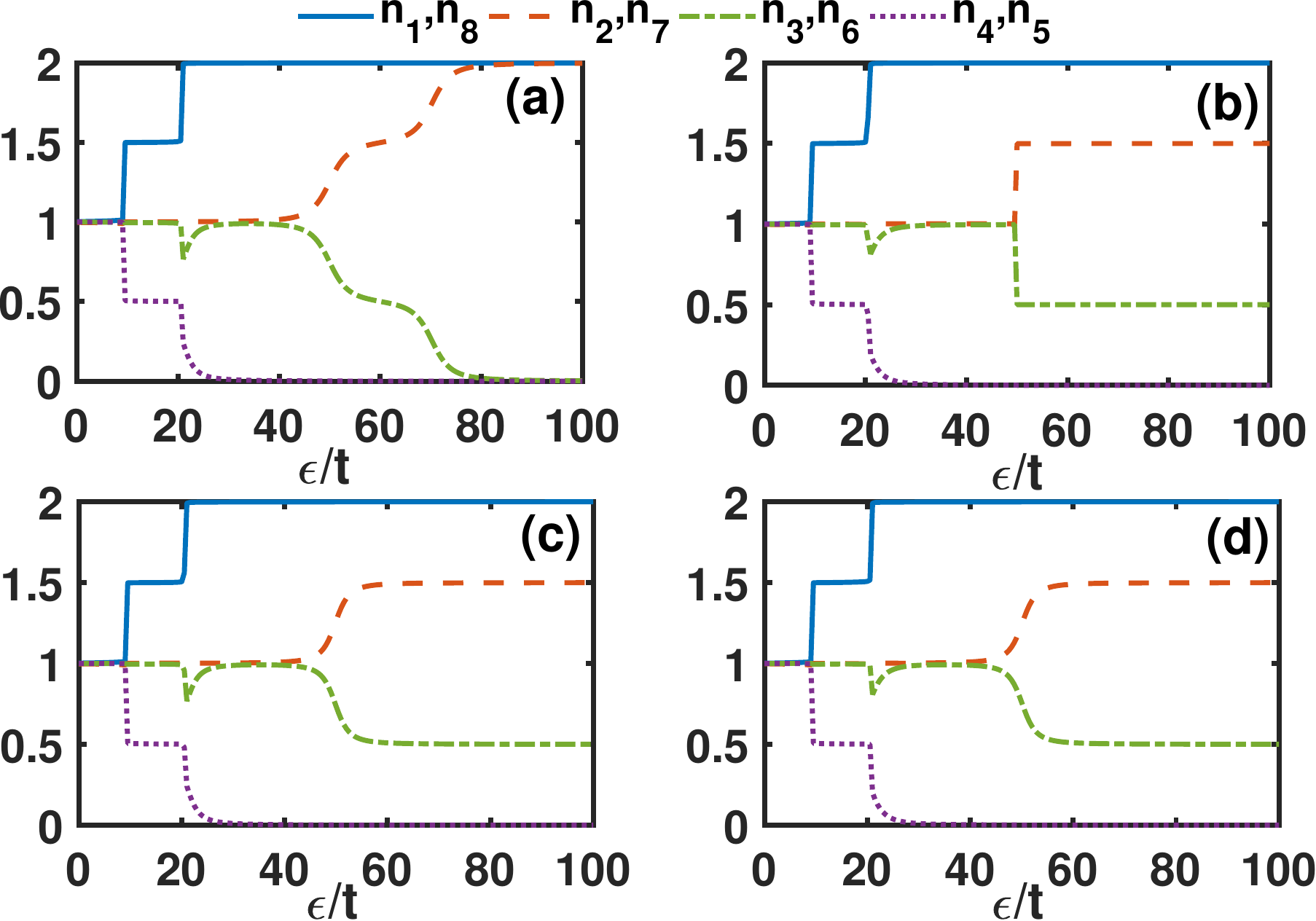}
	\caption{ \textbf{Spin ladder.} Charge occupancies of different sites as a function of $\epsilon/t$ in a spin ladder of size $2 \times 4$ for different eigenstates: (a) $|S_1\rangle$; (b) $|S_2\rangle$;  (c)  $|T_1\rangle$; and (d) $|T_2\rangle$. The ground state $|S_1\rangle$ can be truly certified through charge detection when $\epsilon/t>80$. }
	\label{Fig_Ladder_Nh4_Nv2}
\end{figure} 

\section{Two dimensional lattices} 

In a many-body 1D system with local interactions, \textit{area laws}~\cite{eisert2010colloquium} dictate that the entanglement of a subsystem with the rest of the system in the low-energy spectrum is fixed, with possible logarithmic corrections in critical systems~\cite{calabrese2004entanglement}. This allows for efficient simulations of the low-energy spectrum of 1D systems on classical computers~\cite{schollwock2005density,orus2019tensor}. However, for higher dimensional systems for which the area grows with the size of the subsystem, there is no general efficient approach to simulate even the ground state of a many-body system on a classical computer.  For instance, in Ref.~\cite{lauchli2016s} the simulation of the ground state in a 2D Kagome Heisenberg lattice with 48 spins through implementing all possible symmetries in exact diagonalization puts a limit on the ability of classical computers to simulate 2D many-body systems.
We demonstrate the validity of our protocol in 2D in the context of the certification of the ground state of a spin ladder of size $2 \times 4$, depicted in Fig.~\ref{Fig_Schematic}d), with nearest neighbor interactions of the form given in Eq.~(\ref{Hubbard_Hamiltonian}). Note that the results are valid for general 2D Heisenberg lattices. The same tilting potential $\tilde{\epsilon}_k=\tilde{\epsilon}_{N-k+1}=(k-1)\epsilon$ is applied to the ladder sites $k$ (with $k\leq N/2$) and $N-k+1$ . The results are shown in Fig.~\ref{Fig_Ladder_Nh4_Nv2} with the charge occupancies of different sites for the first four eigenstates of the system, namely $|S_{1,2}\rangle$ and $|T_{1,2}\rangle$, as a function of the tilting potential $\epsilon$. The charge occupancies for $|S_1\rangle$ (see Fig.~\ref{Fig_Ladder_Nh4_Nv2}(a))  are clearly distinct from the other eigenstates when the system is tilted at $\epsilon/t>80$. The other higher energy eigenstates cannot be distinguished as they share similar charge configurations due to extra spatial freedom for electrons to restructure themselves.

In order to understand the charge configurations in two dimensional systems it is very insightful to carefully check the wave function of the simplest spin ladder, namely a spin ring of $N=4$. When $\epsilon/t \gg 1$, the charge wave function of the four eigenstates are as following
\begin{eqnarray} \label{ladder_wave_func}
|S_1\rangle&=& \left|\left( {\begin{array}{cc}
	2 & 0 \\
	2 & 0 \\
	\end{array} } \right)
\right\rangle \cr
|T_1\rangle=|S_2\rangle&=& \left|\left( {\begin{array}{cc}
	1 & 1 \\
	2 & 0 \\
	\end{array} } \right)
\right\rangle - 
\left|\left( {\begin{array}{cc}
	2 & 0 \\
	1 & 1 \\
	\end{array} } \right)
\right\rangle
\cr
|T_2\rangle &=& \left|\left( {\begin{array}{cc}
	1 & 1 \\
	2 & 0 \\
	\end{array} } \right)
\right\rangle + 
\left|\left( {\begin{array}{cc}
	2 & 0 \\
	1 & 1 \\
	\end{array} } \right)
\right\rangle
\end{eqnarray}   
where, $\left|\left( {\begin{array}{cc}
	n_1 & n_2 \\
	n_4 & n_3 \\
	\end{array} } \right) \right\rangle$ 
represents the wave function with charge occupation number $n_k$ at site $k$ and for simplicity we have omitted the normalization factors. While for $|T_{1,2}\rangle$ the electrons at the two sites with occupancies $n_k=1$ are in spin triplet states they take spin singlet for the case of $|S_2\rangle$. This guarantees the the orthogonality of $|T_1\rangle$ and $|S_2\rangle$ through their spin degrees of freedom while both sharing exactly the same charge wave function. The same behavior happens for larger lattices, such as the case of size $2 \times 4$ in the main text, where the ground state $|S_1\rangle$ can be fully discriminated from the other three eigenstates through a single charge detection at a high tilting potential.

\section{Experimental realization} \label{sec:Experimental_Realization}

Among existing platforms for quantum simulations \cite{bernien2017probing,zhang2017observation,dutta2012nonequilibrium,barends2014superconducting,o2016scalable,roushan2017spectroscopic,barends2015digital}, fermionic optical lattices~\cite{schreiber2015observation} and semiconductor systems such as quantum dots \cite{watson2018programmable,hensgens2017quantum,li2018crossbar,zajac2016scalable,nakajima2017robust} and dopant arrays \cite{fuechsle2012single,salfi2016quantum} offer a scalable platform with the natural presence of Fermi statistics (as opposed to simulating fermions with bosonic qubits via non-local interactions), as well as Coulomb, tunnel, electron-phonon and spin-orbit interactions. 
The atomic precision of scanning tunneling microscopy lithography~\cite{fuechsle2012single} provides the required versatility to fabricate 1D or 2D phosphorus donor-bound spin arrays in silicon. Calibrated charge sensors can be defined in the proximity of the donors structure to accurately extract charge configurations of 
single~\cite{mahapatra2011charge,elzerman2004single,pla2012single} and pairs~\cite{petta2005coherent,shulman2012demonstration,broome2017high,gray2016unravelling,banchi2016entanglement,gray2018machine} of electron spins. State-of-the-art charge sensors can successfully measure the charge configuration of up to four quantum dots in an array~\cite{thalineau2012few,dehollain2019nagaoka}. A similar number should be valid for dopant arrays too.
The charging energy of phosphorus dopants in silicon is $U{\sim} 47$ meV and both $t$ and $V$ can be tuned via inter-dopant distances. For dopants placed $10$ nm apart, $t$ is about $1$ meV~\cite{gamble2015multivalley} and $V$ about $10$ meV, as considered throughout this letter. 
We note that these values bring the system in a spin density wave phase, close to a transition to a charge density wave predicted to occur at $U/V=2$~\cite{PhysRevLett.53.2327}, which could be of interest for future work.
From these values, the evolution can be considered as adiabatic if $T_{max} {\ge} 13$ ns. 
Experimental charge dephasing values can be converted to $\gamma {\sim} 0.02-1$ $\mu$eV~\cite{pashkin2003quantum,dupont2013coherent,van2018microwave,van2018time}. 
The ratio $\gamma/t$ is found to be ${\sim} 10^{-5}{-}10^{-3}$, as strong tunneling interactions are considered here. As shown in Fig.~\ref{Fig_Decoherence_ST_N4}(c), this results in $d{>}20$ and fidelities above $0.8$ and hence certification can be achieved in dopant systems.
The hyperfine interactions, coupling electron and nuclear spins in dopant atoms, constitute another possible source of errors in dopant systems as they mix the singlet and triplet subspaces. For the hyperfine coupling of $A{\sim} 0.4$ $\mu$eV, this mixing rate is ${\sim} A^2/(E_{T_1}-E_{S_1})$. 
As the minimum $E_{T_1}{-}E_{S_1} {\sim} 100$ $\mu$eV is found for $N{=}4$ the role of hyperfine mixing rate can be neglected  in comparison with the energy gap $E_{T_1}{-}E_{S_1}$.
However, as the energy gap scales as $1/N^2$, we estimate that hyperfine interactions and thus nuclear spin initialization will become relevant for $N{>}20$.

\section{Large Chains}\label{section_large_chain}

\begin{figure} \centering
	\includegraphics[width=8cm,height=7cm,angle=0]{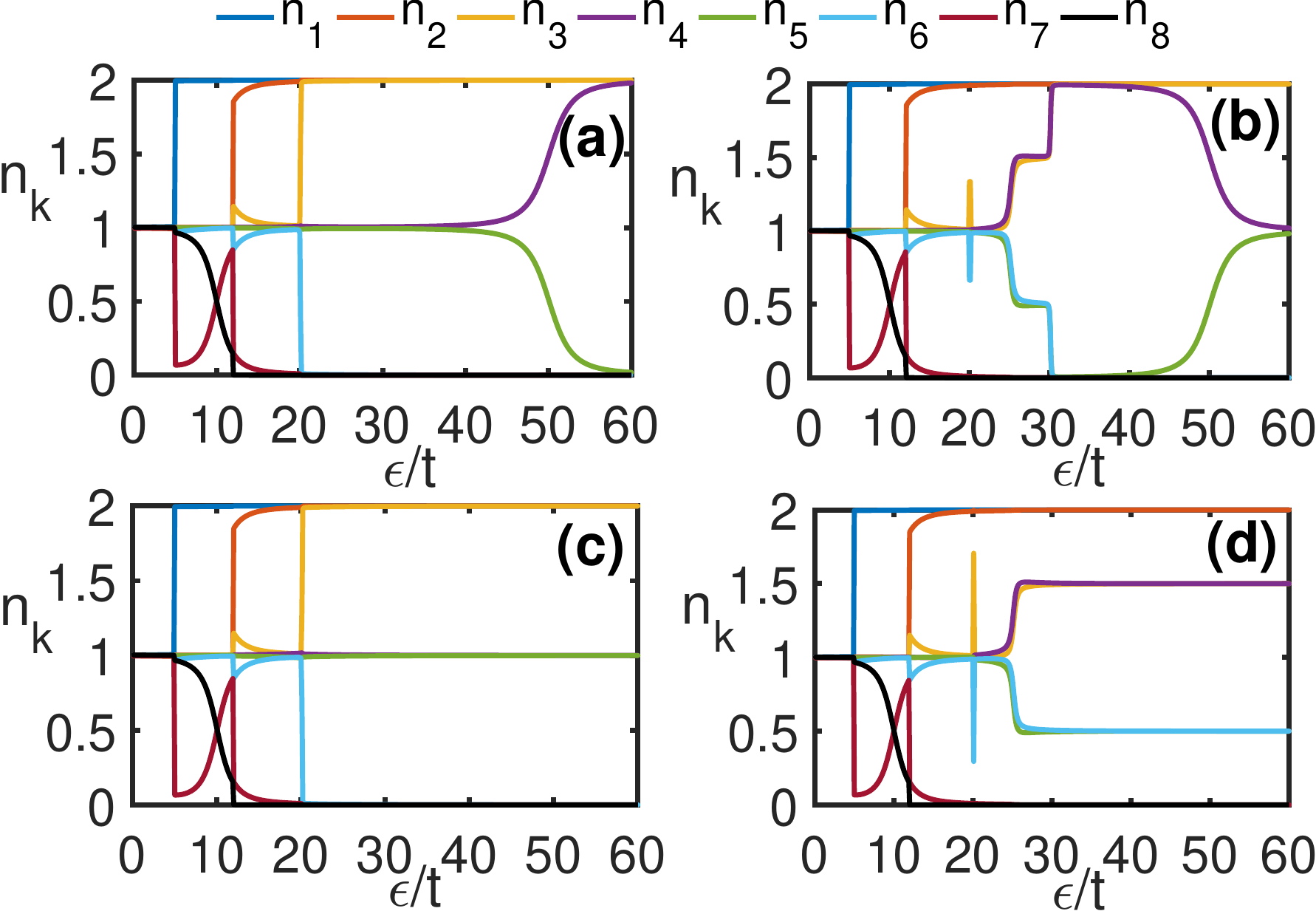}
	\caption{ \textbf{Charge configuration for a chain of size $N=8$.} The charge configuration of a spin simulator with $N=8$,  $U/t=40$ and $V/t=10$. The charge configrations are given for: (a) the ground state $|S_1\rangle$; (b)  the first singlet excited state $|S_2\rangle$; (c) the first triplet excited state $|T_1\rangle$; and (d) the second triplet eigenstate $|T_2\rangle$.  }
	\label{Fig_Supp_Charge_ST_N8}
\end{figure}

\begin{figure*} \centering
	\includegraphics[width=12cm,height=6cm,angle=0]{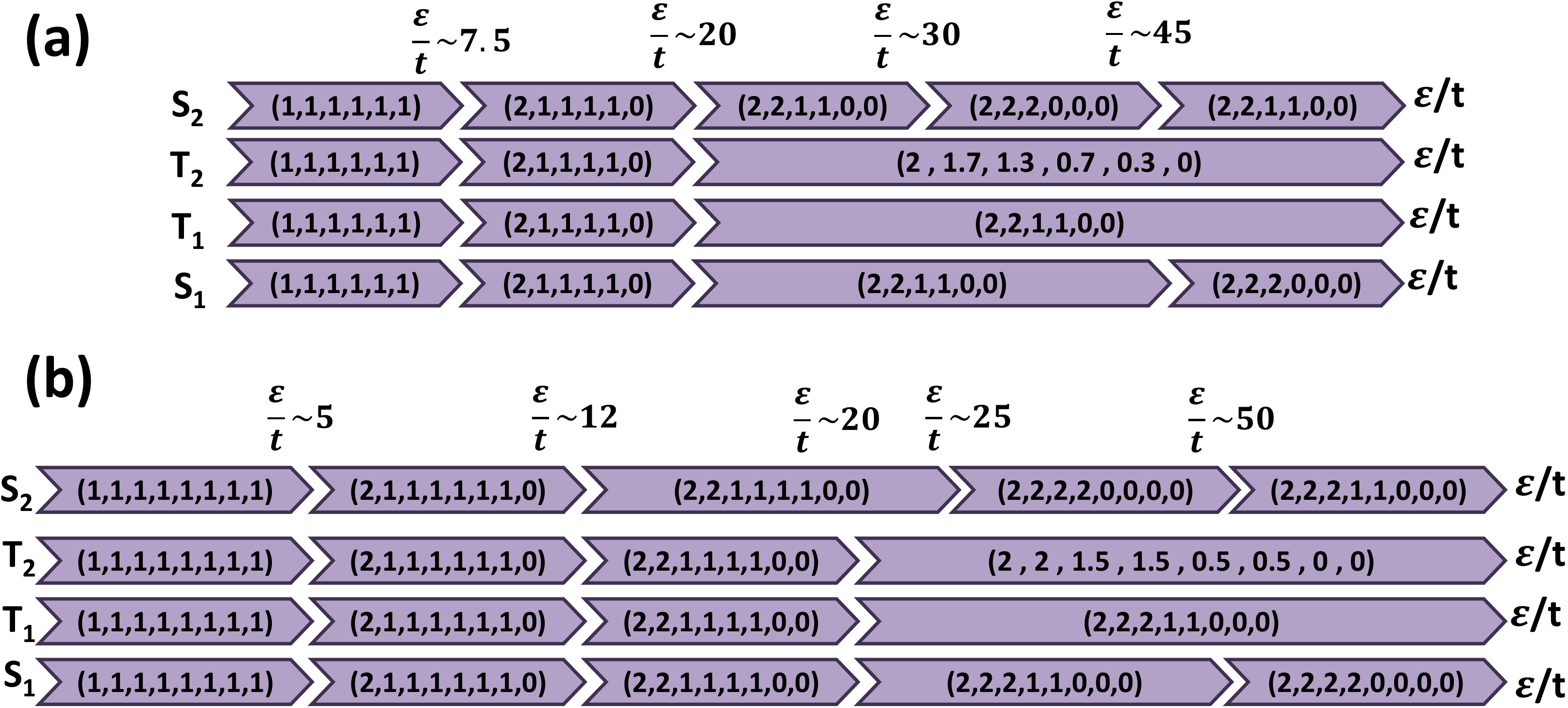}
	\caption{ \textbf{Schematic of charge configurations.} Charge configurations for the first four eigenstates as the tilting potential is varied for: (a) a chain of length $N=6$; and (b) a chain of length $N=8$.  }
	\label{Fig_Supp_Schematic}
\end{figure*}

In this section, we show how our certification protocol scales favorably in the context of systems too large to be tractable classically.
In fact, the proposed mechanism can also be applied to large chains with $N>4$. The charge configurations become more diverse as the size of the system increases. 
We plot in Fig.~\ref{Fig_Supp_Charge_ST_N8} the charge occupancy evolution of the first four eigenstates $|S_{1,2}\rangle$ and $|T_{1,2}\rangle$ for a 8 site-chain, using the same experimental parameters discussed above, namely $U/t=40$ and $V/t=10$.
The overall picture is similar to the $N=4$ case except that there are more charge movements. The eigenstate charge configurations as a function of $\epsilon/t$ for chains of length $N=6$ and $N=8$ are represented schematically in Figs.~\ref{Fig_Supp_Schematic}(a)-(b). It can be shown that the final configuration of the eigenstate $|S_1\rangle$ is always $(2,\cdots,2,0,\cdots,0)$ and for the eigenstate $|T_1\rangle$ it is $(2,\cdots,2,1,1,0,\cdots,0)$.
An important feature which arises in large chains is that the final charge configuration of $|T_2\rangle$ shows partial charge occupancies. This is due to a superposition of charges. 

Remarkably, independently of the system size we can discriminate between the four eigenstates using only two potential tilts. For instance, in the case of $N=6$, with $\epsilon/t{=}35$ we can fully discriminate the eigenstate $|S_2\rangle$ and $|T_2\rangle$ from the rest but we cannot distinguish $|S_1\rangle$ from $|T_1\rangle$. Note that, at this value of the potential tilt the charge measurement outcome for $|T_2\rangle$ is not unique as that eigenstate is a superposition of different charge configurations, but due to orthogonality it does not share any charge configuration with $|T_1\rangle$ (which has the same charge configuration as $|S_1\rangle$) and $|S_2\rangle$. If the charge measurement shows the configuration $(2,2,1,1,0,0)$ this means that the quantum state is either $|S_1\rangle$ or $|T_1\rangle$ and to discriminate between them one has to tilt the system further to $\epsilon/t{=}70$ for which  the two eigenstates take different charge configurations. The same argument is valid for $N=8$ in which the two measurements should also be performed at $\epsilon/t{=}35$ and $\epsilon/t{=}70$ for full discrimination between the four eigenstates. We have also performed the simulation for $N=10$ (data not shown) in which again two measurements at different tilts are enough to fully distinguish the four eigenstates.  

For larger systems, whose ground state might become non-tractable classically and for which the global charge state cannot be measured using a single detector anymore, we propose to divide the chain into several modules. Each module should be certified independently, which results in the total number of measurements to still scale linearly with the system size as well as with the number of eigenstates to discriminate.   
   
\section{Conclusion} 

We have proposed an efficient procedure for certifying the performance of spin-based quantum simulators via discriminating between their low energy eigenstates without using quantum tomography. This task is nontrivial as the eigenstates cannot be distinguished locally due to many-body entanglement and to the eigenstates sharing the same symmetries and total spin numbers. 
Our certification scheme does not require individual spin measurement, but only rely on global charge measurements, thus facilitating practical scale up to large system sizes. We demonstrate how a large system, not classically tractable, can be certified block-by-block, with a number of required measurements which increases only linearly with the number of blocks and eigenstates to discriminate. We identify realistic conditions for the block size, in terms of charge coherence, tilt speed and charge detection, for our certification scheme to be implemented experimentally. Successfully certifying all the blocks of a quantum simulator will maximize the confidence into the outputs given by the whole system in a classically non-tractable regime. After certification of the spin Hamiltonian in the low energy regime, the same simulator could potentially be used to reveal classically inaccessible features such as long-time dynamics and complex 2D structures.\\

\textbf{Acknowledgments}. The authors would like to thank Didier ST Medar for helpful discussions. AB thanks the National Key R\&D Program of China, Grant No. 2018YFA0306703. We also acknowledge support from​ the ARC Centre of Excellence for Quantum Computation ​and Communication Technology (CE170100012)​, Silicon Quantum Computing Pty. Limited and an ARC Discovery Project (DP180102620). JS acknowledges support from an ARC DECRA fellowship (DE160101490). SB and AB acknowledge support from the EPSRC Non-Ergodic Quantum Manipulation program grant EP/R029075/1.


\end{document}